\documentclass[aps,pra,floats,twocolumn,twoside,floatfix]{revtex4}
\usepackage{amsmath}
\usepackage{graphicx}

\begin{document}


\title{Multiple solutions of coupled-cluster equations for PPP model of  [10]annulene}
\author{Rafa{\l}  Podeszwa and Leszek Z. Stolarczyk} 
\affiliation{Department of Chemistry, University of Warsaw, Pasteura 1, PL-02-093 Warsaw, Poland}

\begin{abstract}

Multiple (real) solutions of the CC equations (corresponding to the CCD, ACP and ACPQ methods)
are studied for the PPP model of [10]annulene, ${\rm C}_{10}{\rm H}_{10}$. 
The long-range electrostatic  interactions are represented either by 
the Mataga--Nishimoto potential, or Pople's $R^{-1}$ potential.  
The multiple solutions are obtained in a quasi-random manner, by generating 
a pool of starting amplitudes and applying a standard CC iterative procedure 
combined with Pulay's DIIS method. Several unexpected features of these 
solutions are uncovered, including the switching between two CCD solutions 
when moving between the weakly and strongly correlated regime of the PPP 
model with Pople's potential.

\end{abstract}

\maketitle



\section{Introduction}

The coupled-cluster (CC) method~\cite{cizek:66} is one of the basic tools for studying 
the electronic-correlation effects in many-electron systems.  Due to their nonlinearity, 
the CC equations may have multiple solutions; this problem was first analyzed by 
\v{Z}ivkovi\'{c} and Monkhorst~\cite{zivkovic:78}. Recently,  Kowalski and Jankowski 
~\cite{kowalski:98a} applied to the CC equations a new powerful 
mathematical tool---the homotopy (continuation) method~\cite{drexler:78}. 
By using this tool they were able to find  the complete sets of solutions for the CCD and 
the CCSD equations corresponding to some four-electron four-orbital systems, known as 
the H4 and P4 models. Very recently, Podeszwa {\it et al.\/}~\cite{podeszwa:02b} used 
the homotopy method in a study of multiple solutions to the CCD equations for the  
Pariser-Parr-Pople (PPP) model~\cite{pariser:53a,pople:53} of benzene (6 electrons, 
6 orbitals). The available data on the CC multiple-solution problem is still very 
limited. The purpose of this paper is to get some insight into the multiple solutions of the 
CCD equations (as well as the equations of some modified CCD methods) for the PPP 
model of [10]annulene (10 electrons, 10 orbitals), a system which is already too big to be 
treated by the homotopy method. 

Benzene and [10]annulene are the first two members of the family of cyclic polyenes
(annulenes) ${\rm C}_M {\rm H}_M$,  where $M=4m_0+2$, $m_0 = 1, 2, \ldots$,
and the assumed point-symmetry group is D$_{M{\rm h}}$.  
The problem of electronic correlations in annulenes, which may serve as 
prototypes of one-dimensional metal,  have been thoroughly studied within the PPP model 
by Paldus and co-workers~\cite{paldus:84a,paldus:84b,paldus:84c}. 
Due to the high symmetry of annulenes, the contributions from the single 
excitations are absent in the PPP model, and the CCSD method reduces to the CCD one. 
Paldus {\it et al.\/} discovered and 
documented the breakdown of the CCD method for these systems. 
They also showed that this breakdown may be attributed to the neglect 
in the CCD method the terms providing the coupling between the D and Q excitations. 
Paldus {\it et al.\/} devised an approximate coupled-pair method, 
called ACPQ~\cite{paldus:84a,paldus:84b}, in which these coupling terms (approximately) 
canceled certain quadratic terms in the CCD equations. Unlike the CCD method, 
the ACPQ method, and a related ACP-D45 method (ACP in short) 
introduced earlier~\cite{jankowski:80}, were shown to perform well for the annulenes, 
being convergent and giving the correlation energies close to 
the full configuration-interaction (FCI) results. However, a recent CC study~\cite{podeszwa:02a} 
of annulenes, taking into account the double (D), triple (T), and quadruple (Q) 
excitations in the CC operator, showed that even the CCDTQ method breaks down 
for these systems when the correlation effects become sufficiently strong. 
Moreover, the ACP and ACPQ methods were shown to fail for very large 
annulenes~\cite{podeszwa:02c}. Thus, the annulenes remain 
a challenging problem for the many-electron theories.

\section{Finding multiple solutions of CCD, ACP, and ACPQ equations for [10]annulene}

In the CCD method, the CC operator corresponding to the double excitations depends 
on some linear parameters, hereafter referred to as the $t_2$ amplitudes. The CCD 
equations comprise a set of quadratic equations for the unknown $t_2$ amplitudes. 
In the nonorthogonally spin-adapted formalism (see, e.g., Ref.~\onlinecite{stolarczyk:84}), 
the number of $t_2$ amplitudes is equal 
to $K = n_{\rm o} n_{\rm u} (n_{\rm o} n_{\rm u} + 1)/2$, where $n_{\rm o}$ and $n_{\rm u}$ 
is the number of occupied and unoccupied orbitals, respectively 
(no symmetry other then the spin one is assumed). According to the B{\'e}zout theorem,
such a set of quadratic equations may have up to $2^K$ solutions,
a truly astronomical number in most applications. Therefore, finding the complete set 
of CCD solutions may be attempted only when the number of $t_2$ amplitudes is very small 
(e.g., due to the spatial symmetry of the many-electron system). 
In the homotopy study of the H4 model~\cite{kowalski:98a} 
(corresponding to 6 spin- and spatial-symmetry--adapted $t_2$ amplitudes), 
the complete set of CCD solutions numbered 12, which is distinctly smaller than 
the B{\'e}zout limit (64). On the other hand, the application of the homotopy method 
to the PPP model of benzene~\cite{podeszwa:02b} (having 8 $t_2$ amplitudes adapted to spin, 
spatial, and time-reversal symmetry) has brought in some quite disturbing findings:
not only the total number of solutions (230) approached the theoretical limit (256),
but some solutions led to the correlation energies much more negative than that 
of the ``standard'' CCD solution (which approximate the ground-state FCI solution); 
in addition all the ``nonstandard'' solutions were hardly similar to the FCI ones.

In the PPP model of [10]annulene, the CCD equations adapted to spin, spatial, 
and time-reversal symmetry correspond to 29 $t_2$ amplitudes. The  ACP and ACPQ equations 
are obtained by neglecting (ACP and ACPQ) and modifying (ACPQ) some of the quadratic 
terms in the CCD equations, see, e.g., Ref.~\onlinecite{podeszwa:02c}.  
In the present paper we use the PPP model of annulenes described in Ref.~\onlinecite{podeszwa:02a} 
and follow the notation used there: $\beta$ $( \le 0)$~eV is the resonance 
integral of the PPP model, and $\gamma(R)$ is a potential describing the long-range 
electrostatic interactions. We shall report the results obtained for 
two variants of the PPP model: the PPP-MN model, 
employing the Mataga-Nishimoto potential~\cite{mataga:57} 
($\gamma(R) =e^2[R + e^2 (\gamma^0)^{-1}]^{-1}$), with $\gamma^0 = 10.84$~eV, 
and the PPP-P model, using Pople's potential~\cite{pople:53} ($\gamma(R) =e^2/R$ for $R \neq 0$),
with $\gamma^0 = 14.25878$~eV ($e$ is the elementary charge).

If the CCD results for benzene~\cite{podeszwa:02b} may be extrapolated, 
the CCD (and ACP and ACPQ) equations for [10]annulene may have more than $10^8$ solutions;
that makes the application of the homotopy method hopeless in this case. 
On the other hand, it was found  in Ref.~\onlinecite{podeszwa:02b} that some of the real 
CCD solutions for benzene were stable within their close neighborhoods: such a solution 
(subject to the numerical rounding errors) led to a convergent iteration process when the 
direct-inversion-of-the-iterative-subspace (DIIS) method~\cite{pulay:82} was used. 
Thus, one may expect that the DIIS method is able to trace various solutions, 
if only the starting points are suitably chosen.  

A standard method of solving the CCD equations is based on the iterative procedure 
represented in Eq.~(14) of Ref.~\onlinecite{podeszwa:02b}. 
By starting from the $t_2$ amplitudes equal to zero, one recovers in the first iteration 
the $t_2^{(1)}$ amplitudes corresponding to the first order of the M{\o}ller-Plesset (MP) 
perturbation theory; in the subsequent iterations higher-order MP contributions are added. 
Such an iteration procedure, if convergent, furnishes the standard CCD solution corresponding 
(as it is often {\it a priori\/} assumed) to the ground state of the system. 
The CCD electronic-correlation energy is then calculated as a certain linear 
function of the converged $t_2$ amplitudes (the $t_2^{(1)}$ amplitudes correspond 
to the second-order MP correlation energy and will be, after Ref.~\onlinecite{podeszwa:02a}, 
referred to as the MP2 amplitudes). The above treatment applies also to the ACP and ACPQ 
equations. In annulenes one may also use a kind of  analytical-continuation procedure 
to find the standard solution for those values of parameter $\beta$ for which the standard 
procedure (starting from the MP2 amplitudes) does not converge:
one uses as the starting point the converged $t_2$ amplitudes found 
for the sufficiently close $\beta$ value.

The multiple solutions of the CCD, ACP, and ACPQ equations presented in this paper were 
found in a quasi-random manner, by generating a pool of the initial $t_2$ amplitudes 
and then applying the DIIS procedure. The starting pool included the results of several 
hundred  converged CCD, ACP, and ACPQ calculations obtained for different $\beta$ values 
and for different forms of function $\gamma(R)$ (these sets of $t_2$ amplitudes 
indeed looked random).  Most of the calculations starting from this quasi-random 
amplitudes either diverge, or converge to the standard solution, but in some cases a new 
solution is obtained. Certainly, this is not a very efficient method of finding the 
multiple solutions, but since a single CCD calculation 
for [10]annulene usually takes only several seconds, a lot of data may be generated 
that way. We have found that the strongly correlated regime ($\beta > -1.5$~eV) 
is rich in multiple solutions, while in the weakly correlated regime ($\beta < -3.0$~eV) 
practically only the standard solution could be reached that way. Therefore, 
we performed calculations for several values of $\beta$ 
from interval $(-1.0\mbox{ eV}, 0 \mbox{ eV})$. These results were then extended 
as far as possible by a careful application of our analytical-continuation procedure. 
The obtained results are not supposed to provide the complete set of real solutions 
for [10]annulene. Nevertheless, they provide a rather intriguing glimpse at 
the ``tip of the iceberg'' of the multiple solution problem for this system. 
Let us note that the multiple solutions of the ACP and ACPQ equations have never been studied before.

\section{Results}

\begin{figure}[thb]
\caption{PPP-MN model of [10]annulene, CCD equations. Correlation energies corresponding 
to multiple solutions; ground-state FCI results shown for comparison.}
\begin{center}
\includegraphics[scale=1.0]{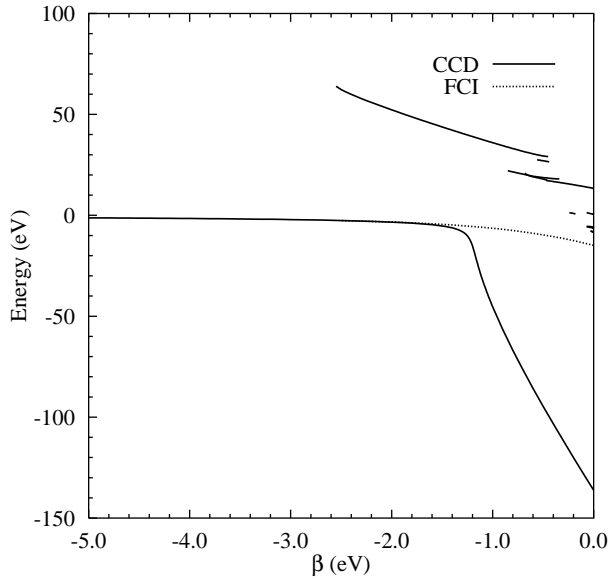}
\end{center}
\label{fig:mn_ccd}
\end{figure}

\begin{figure}[thb]
\caption{PPP-MN model of [10]annulene, ACP equations (compare Fig.~\ref{fig:mn_ccd}).}
\begin{center}
\includegraphics[scale=1.0]{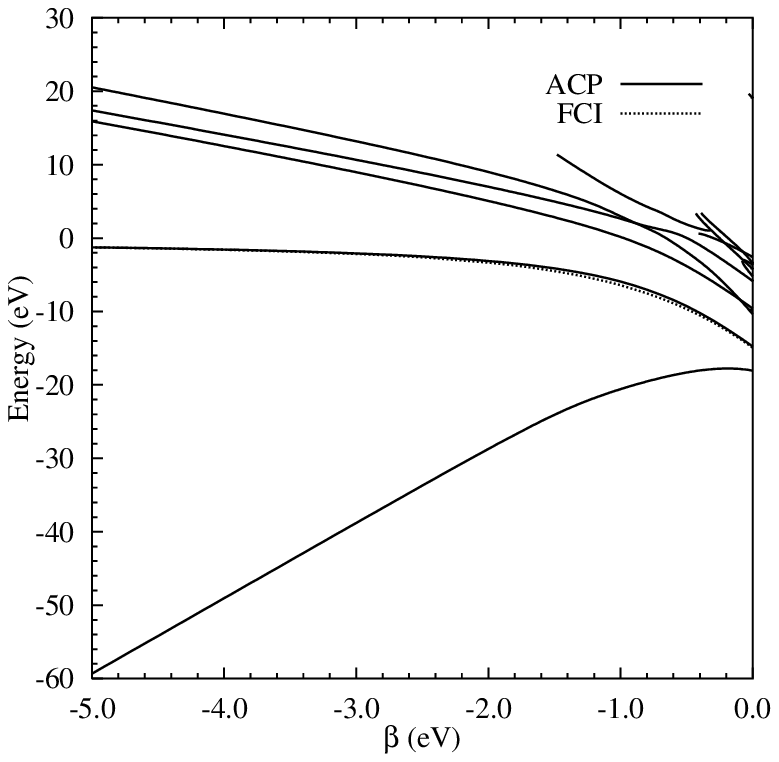}
\end{center}
\label{fig:mn_acp}
\end{figure}

\begin{figure}[thb]
\caption{PPP-MN model of [10]annulene, ACPQ equations (compare Fig.~\ref{fig:mn_ccd}).}
\begin{center}
\includegraphics[scale=1.0]{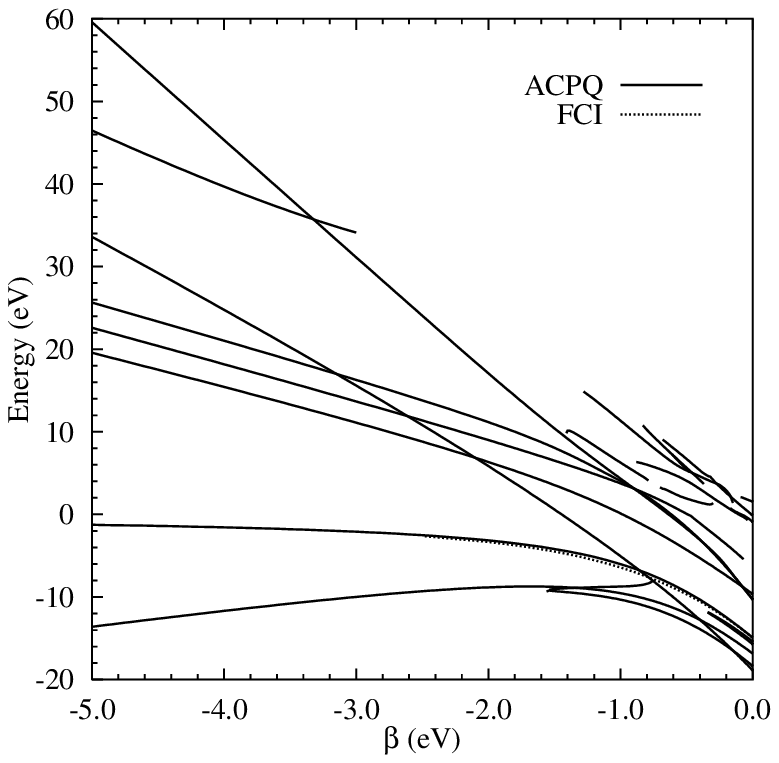}
\end{center}
\label{fig:mn_acpq}
\end{figure}

In Figs.~\ref{fig:mn_ccd}--\ref{fig:mn_acpq} we present the 
electronic-correlation energies corresponding to the multiple solutions of the CCD, ACP, 
and ACPQ equations, respectively. These results are obtained within the PPP-MN model 
for $-5.0 \mbox{ eV} \le \beta < 0 \mbox{ eV}$, each solution is represented by a continuous line. 
The line endpoints (not touching the graph's boundaries) indicate the limiting $\beta$ 
values corresponding to our analytical-continuation procedure. It is seen that some of the 
new solutions exist for a very broad range of $\beta$, while some can be continued only 
within a very narrow interval. The ground-state FCI correlation energies, shown as the 
broken line, are provided for a reference. The solution line that approaches the FCI line 
in the weakly correlated regime corresponds to the standard solution discussed in the 
previous papers~\cite{paldus:84a,podeszwa:02a}.

\begin{figure}
\caption{PPP-P model of [10]annulene, CCD equations (compare Fig.~\ref{fig:mn_ccd}).}
\begin{center}
\includegraphics[scale=1.0]{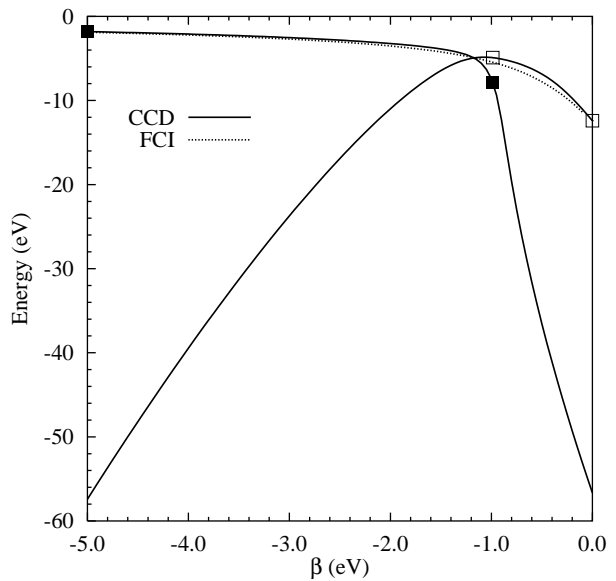}
\end{center}
\label{fig:ppp-p_ccd}
\end{figure}

The CCD results in Fig.~\ref{fig:mn_ccd} are rather simple: The standard-solution line
is the only extending for all the $\beta$ values. (When the standard iteration procedure 
is combined with the DIIS technique, the standard CCD solution for [10]annulene  
can be obtained for all negative values of $\beta$, as shown previously by 
Paldus {\it et al.}~\cite{paldus:84a} by using the Newton-Raphson method.) 
In the strongly correlated regime ($\beta > -1.5$~eV), the standard CCD solution 
provides a rather poor approximation to the ground state, and the corresponding 
correlation energy falls well below the FCI line. Other solutions have the energies 
higher than the FCI result, some of them may represent excited states. 
Most of them can be found only in the proximity of $\beta=0$, where the 
quasidegeneracy effects are the strongest.

\begin{table*}[t]
\caption{PPP-P model of [10]annulene, CCD equations. Correlation energies for 
two solutions, CCD1 and CCD2; ground-state MP2 and  FCI results shown for comparison 
(all energies in eV). $\theta$ (in deg) and $\eta$ gauge the similarity of a given solution 
to the FCI solution (see text).}
\label{tab:ccd1-ccd2}
\begin{ruledtabular}
\begin{tabular}{rrrrrrrrrr}
& \multicolumn{3}{c}{$\beta=-2.500$} & \multicolumn{3}{c}{$\beta=-1.175$} &
 \multicolumn{3}{c}{$\beta=-0.500$} \\
& \multicolumn{1}{c}{$E_{\rm corr}$} &\multicolumn{1}{c}{$\theta$} &
\multicolumn{1}{c}{$\eta$} & \multicolumn{1}{c}{$E_{\rm corr}$} & \multicolumn{1}{c}{$\theta$} & \multicolumn{1}{c}{$\eta$} &
\multicolumn{1}{c}{$E_{\rm corr}$} & \multicolumn{1}{c}{$\theta$} & \multicolumn{1}{c}{$\eta$}\\
\hline
CCD1 & $  -2.825$ & $   5.1$ & $  0.92$ & $ -4.937$ & $19.0$ & $1.30$ & 
$-36.539$ & $46.4$ & $8.05$ \\ 
CCD2 & $-16.896$ & $ 91.3$ & $15.00$ & $ -4.937$ & $88.2$ & $1.14$ & 
$  -6.856$ & $50.9$ & $0.53$ \\ 
MP2   & $  -1.933$ & $ 23.1$ & $  0.66$ & $ -2.388$ & $42.9$ & $0.41$ & 
$  -2.714$ & $57.3$ & $0.20$ \\ 
FCI     & $  -3.045$ & $   0.00$ & $  1.00$ &$ -4.908$ & $  0.00$ & $1.00$ & 
$  -7.769$ & $  0.00$ & $1.00$ \\ 
\end{tabular}
\end{ruledtabular}
\end{table*}

The ACP and ACPQ results (Figs.~\ref{fig:mn_acp}  and ~\ref{fig:mn_acpq} 
respectively) display a different and more complicated behavior. Unlike the CCD 
standard solution, the ACP and ACPQ standard solutions stay close to 
the FCI line also in the strongly correlated regime (although the corresponding $t_2$ 
amplitudes show increasing deviations from the FCI $t_2$ amplitudes as $\beta 
\rightarrow 0$~\cite{podeszwa:02a}). The ACP and ACPQ standard solutions 
can be found for $\beta < 0$ by applying the standard iteration procedure 
starting from the MP2 amplitudes~\cite{podeszwa:02a}. Quite a few nonstandard solutions have been 
found, some of them extending for the whole range of $\beta$.  Quite unusual is the 
presence of the solution lines lying below the standard-solution line 
(the ``underground'' lines). Such exotic solutions have been observed 
in the study of benzene~\cite{podeszwa:02b}; however, they were unstable 
in the DIIS iterative procedure (could be found only by the homotopy method). 
The ``underground'' solutions  obviously have no physical meaning, and it is a 
little disturbing that they may appear while iteratively solving the ACP or ACPQ equations. 
One of the ACPQ ``underground-solution'' lines is especially peculiar: it 
starts with the lowest energy for $\beta=0$ and rises quite steeply with 
the decreasing $\beta$, crossing several solution lines, including the standard-solution 
line (apparently there is no ``non-crossing rule'' for different solution lines).

Quite intriguing are the solutions of the CCD equations corresponding to the PPP-P 
model, displayed in Fig.~\ref{fig:ppp-p_ccd} together with the ground-state FCI results. 
We were able to find only two solutions, hereafter referred to as CCD1 (the standard solution) 
and CCD2 (the nonstandard one). Despite the use of the extensive quasi-random search, 
the CCD iterations invariably led to CCD1 or CCD2 (these solutions were so strong 
attractors that the iterations rarely diverged). In contrast to the CCD results for 
the PPP-MN model, the iterations starting directly from the MP2 amplitudes converge 
to the CCD1 solution in the weakly correlated regime, while in the strongly correlated 
regime the CCD2 solution is obtained. In Fig.~\ref{fig:ppp-p_ccd} we mark by the squares 
the endpoints of the respective direct-convergence domains. One may thus obtain the CCD1 
and CCD2 solutions in a deterministic way, by starting the calculations within these 
regions. Moreover, by applying our analytical-continuation procedure, we are able to 
extend a given solution line outside its domain of (direct) convergence.

The CCD1-solution line resembles closely the line corresponding to the standard CCD 
solution for the PPP-MN model,  see Fig.~\ref{fig:mn_ccd}, while  the CCD2 line has 
no counterpart in that figure. The CCD1 and CCD2 lines cross for $\beta = -1.17479$~eV,
in the vicinity of the FCI-solution line. In Table~\ref{tab:ccd1-ccd2} we present the 
CCD1, CCD2, MP2 and FCI correlation energies calculated for three representative $\beta$ 
values, including the intersection coordinate. In addition, we show there the values 
of parameters $\theta$ and $\eta$ which provide a measure of similarity between the 
vector of $t_2$ amplitudes corresponding to a given solution and the vector of $t_2$ 
amplitudes extracted from the linear coefficients of the ground-state FCI wavefunction:
$\theta$ is the angle between the vectors, and $\eta$ is the ratio of the vector 
lengths (for details, see Ref.~\onlinecite{podeszwa:02a}).

From  Fig.~\ref{fig:ppp-p_ccd}  and Table~\ref{tab:ccd1-ccd2} a consistent picture 
emerges: The CCD1 solution provides an acceptable approximation to the ground-state 
FCI solution up to the vicinity of the CCD1/CCD2 crossing; then, in the strongly-correlated
region, it deteriorates and hardly deserves the name of the standard solution. The CCD2 
solution, on the other hand, starts as the ``underground'' in the weak correlation 
regime (bearing no resemblance to the FCI solution), but it improves in the 
strongly-correlated regime, and, finally, becomes a slightly better (but still poor)
approximation to the FCI solution (the CCD2 correlation energies are surprisingly similar 
to the FCI ones in the strongly correlated regime). Thus, it looks as  CCD1 and CCD2
swap their roles as the standard and ``underground'' solutions. The behavior of the MP2
results agrees with the observed convergence pattern for iterations starting from the 
MP2 amplitudes. Interestingly, if one performed a  CCD study of the PPP-P model of 
[10]annulene by applying a standard iteration procedure (starting with the MP2 
amplitudes) at some representative $\beta$ values (say, $-5.0$, $-2.5$, $-1.5$, and $-0.5$~eV),
the conspicuous switch between the CCD1 and CCD2 solutions would have been passed unnoticed. 
By looking at the CCD correlation energies  alone, one would then proclaimed a very good 
performance of the CCD method up to very small $\beta$ values, contrary to the results 
found by  Paldus {\it et al.\/}~\cite{paldus:84a} for the PPP-MN model.

The study of the ACP and ACPQ equations in the PPP-P model brings in some additional 
unexpected findings: the standard ACP solution cannot be found for $\beta$ between 
$-0.80$ and $-0.4$~eV, while the standard ACPQ solution encounters problems already 
for $\beta > -2.53$~eV. Apparently, in the strongly correlated regime of the PPP-P 
model both methods become unstable.

Our search of the multiple real solutions of the CCD, ACP, and  ACPQ equations has 
brought  several unexpected findings: (i) while solving the CC equations by 
iterations, one may arrive at an ``underground'' solution, (ii) in the PPP model,
the character of multiple solutions is sensitive to the form of the long-range 
potential (the $\gamma$ function), (iii) different solutions may play the role 
of the standard solution when the strength of the electronic-correlation effects
is varied (this may apply to the changes in molecular geometry, e.g., to the bond
breaking). In conclusion: it seems that the problem of multiple solutions of the 
CC equations deserves still more attention. Some variants of our random-search 
approach may be tried also on larger many-electron systems.

This work was supported in part by the Committee for Scientific Research (KBN) 
through grant No. 7 T09A 019 20.


\begin{thebibliography}{10}
\expandafter\ifx\csname url\endcsname\relax
  \def\url#1{\texttt{#1}}\fi
\expandafter\ifx\csname urlprefix\endcsname\relax\def\urlprefix{URL }\fi

\bibitem{cizek:66}
J.~{{\v C}{\' \i}{\v{z}ek}}, J. Chem. Phys. 45 (1966) 4256.

\bibitem{zivkovic:78}
T.~P. {{\v Z}ivkovi{\' c}}, H.~J. Monkhorst, J. Math. Phys. 19 (1978) 1007.

\bibitem{kowalski:98a}
K.~Kowalski, K.~Jankowski, Phys. Rev. Lett. 81 (1998) 1195.

\bibitem{drexler:78}
F.~J. Drexler, Continuation methods, Academic Press, New York, 1978, p.~69.

\bibitem{podeszwa:02b}
R.~Podeszwa, L.~Z. Stolarczyk, K.~Jankowski, K.~Rubiniec,
  physics/0207086, submitted to J. Chem. Phys.

\bibitem{pariser:53a}
R.~Pariser, R.~Parr, J. Chem. Phys. 21 (1953) 466.

\bibitem{pople:53}
J.~A. Pople, Trans. Faraday Soc. 49 (1953) 1375.

\bibitem{paldus:84a}
J.~Paldus, M.~Takahashi, R.~W.~H. Cho, Phys. Rev. B 30 (1984) 4267.

\bibitem{paldus:84b}
J.~Paldus, J.~{{\v C}{\' \i}{\v{z}ek}}, M.~Takahashi, Phys. Rev. A 30 (1984)
  2193.

\bibitem{paldus:84c}
J.~Paldus, M.~Takahashi, R.~W.~H. Cho, Int. J. Quantum Chem. (Quantum Chem.
  Symposium) 18 (1984) 237.

\bibitem{jankowski:80}
K.~Jankowski, J.~Paldus, Int. J. Quantum Chem. 18 (1980) 1243.

\bibitem{podeszwa:02a}
R.~Podeszwa, S.~A. Kucharski, L.~Z. Stolarczyk, J. Chem. Phys. 116 (2002)
  480.

\bibitem{podeszwa:02c}
R.~Podeszwa, physics/0208076, submitted to Chem. Phys. Lett.

\bibitem{stolarczyk:84}
L.~Z. Stolarczyk, H.~J. Monkhorst, Int. J. Quantum Chem. (Quantum Chem.
  Symposium) 18 (1984) 267.

\bibitem{mataga:57}
N.~Mataga, K.~Nishimoto, Z. Phys. Chem. (Frankfurt am Main) 13 (1957) 140.

\bibitem{pulay:82}
P.~Pulay, J. Comp. Chem. 3 (1982) 556.

\end{thebibliography}
\end{document}